\documentclass{llncs}

\usepackage[T1]{fontenc}
\usepackage[utf8]{inputenc}

\usepackage[usenames,dvipsnames]{color}  
\usepackage[pdfborder={0 0 0},colorlinks=true,urlcolor=blue,linkcolor=blue,citecolor=blue,bookmarks=true]{hyperref}

\usepackage{graphicx}
\usepackage[shortlabels]{enumitem}
\usepackage{amsmath}
\usepackage{amssymb}
\usepackage{amsfonts}
\usepackage{mathtools}
\usepackage{bm}
\usepackage{booktabs}      
\usepackage{multirow}      

\usepackage{newtxtext}
\usepackage{newtxmath}
\usepackage{dsfont}

\usepackage{inconsolata}   

\usepackage{microtype}
\DisableLigatures[f,i,j,F,I,J]{encoding = *, family = *}

\usepackage[british]{babel}
\usepackage{hyphenat}
\hypersetup{pdftitle={Fast GPU Linear Algebra via Compile Time Expression Fusion}}
\hypersetup{pdfauthor={Ryan R. Curtin, Marcus Edel, Conrad Sanderson}}

\begin{document}

\titlerunning{~}
\authorrunning{~}
\pagestyle{empty}

\title{\scalebox{0.89}{Fast GPU Linear Algebra via Compile Time Expression Fusion}}

\author
  {
  Ryan R. Curtin, 
  Marcus Edel\textsuperscript{{\tiny~}$\dagger$},
  Conrad Sanderson\textsuperscript{{\tiny~}$\ast\diamond$}
  }

\institute
  {
  \textsuperscript{$\dagger$}{\tiny~}\textit{Collabora, Canada;}~
  \textsuperscript{$\ast$}{\tiny~}\textit{CSIRO, Australia;}~
  \textsuperscript{$\diamond$}{\tiny~}\textit{Griffith University, Australia}
  }

\maketitle

\begin{abstract}
We describe the Bandicoot GPU linear algebra toolkit for C++,
which prioritises ease of use without compromising efficiency.
Bandicoot's API aims for compatibility with the popular Armadillo CPU linear algebra library,
enabling easy transition for existing CPU-based codebases.
Unlike other GPU-focused toolkits,
Bandicoot uses template metaprogramming
to generate fused GPU kernels directly at \textit{compile-time},
yielding efficient kernels that can saturate memory bandwidth.
This removes the need for run-time overhead or JIT infrastructure.
Empirical results show that Bandicoot outperforms
(sometimes by considerable margins)
commonly-used linear algebra toolkits including PyTorch, TensorFlow, and JAX.
\end{abstract}

%
%
%

\section{Introduction}
\label{sec:introduction}

As workload sizes increase, modern scientific computing and numerical linear algebra have
increasingly come to depend on the use of
graphics accelerators / graphics processing units (GPUs) to accelerate computation.
The use of GPUs is widespread,
most notably in the field of machine learning~\cite{Sze_2017}
where the use of GPUs for inference in large language models is all but
required~\cite{Song_2024},
though GPUs are also used in a wide variety of other fields,
including astrophysics~\cite{bard2013cosmological},
computational fluid dynamics~\cite{niemeyer2014recent},
weather simulations~\cite{michalakes2008gpu}
and
healthcare~\cite{vamathevan2019applications}.

GPUs are built on the idea of {\em single-instruction-multiple-threads} (SIMT)
and thus are particularly well-suited for linear algebra tasks,
where identical or similar computations are often performed over many matrix elements.
For instance, embarrassingly parallel linear algebra operations like the element-wise Schur product $Z = X \odot Y$
can be performed trivially,
with one thread assigned to compute each element of $Z$.
In situations like this,
GPUs can produce speedups of over 100x compared to a CPU-based implementation.

\begin{figure}[t!]
\centering
\includegraphics[width=0.95\textwidth]{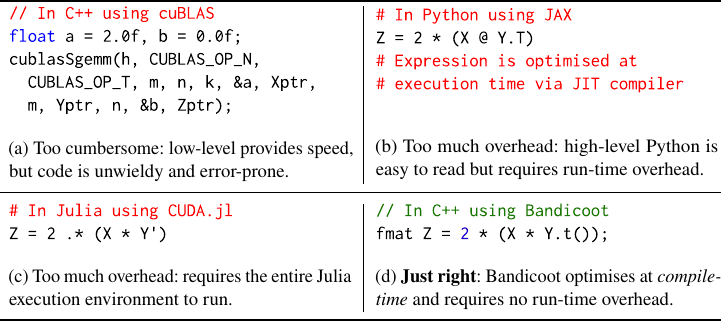}
\vspace*{-2ex}
\caption{Implementations of the simple expression $\mathbf{Z} \gets 2\mathbf{X}\mathbf{Y}^T$ using various toolkits.}
\label{fig:various_toolkits}
\end{figure}

For a practitioner,
there are many options for implementing GPU-accelerated linear algebra
or adapting existing code to run on GPUs (Fig.~\ref{fig:various_toolkits}).
At the low level, vendor-specific toolkits such as CUDA
or cuBLAS can be used,
with each vendor typically supplying its own device-family-specific offering
(e.g. HIP/ROCm, OneAPI, etc).
However, as vendors repeatedly demonstrate little interest in interoperability with competitors' products,
the use of these tools ties the implementation to a particular vendor's GPUs.
Additionally, most practitioners are not comfortable working at such a low level,
which results in significantly increased development effort
that often produces cumbersome code.

On the other end of the spectrum are high-level tools that
aim to entirely abstract the GPU away,
allowing practitioners to focus entirely on their computational tasks.
Tools like MATLAB and Julia~\cite{Bezanson_2017}
aim to provide simple APIs so that practitioners can write visually simple code
that reflects their mathematical intentions.
In the field of machine learning, commonly-used machine learning toolkits like
PyTorch~\cite{Ansel_2024} and
TensorFlow~\cite{Abadi_2016}
are essentially GPU-based linear algebra engines.
Unfortunately, such high-level languages typically require environments
that add considerable overhead at run-time, including increased execution time and resource usage.
Furthermore, sometimes add-ons
like \mbox{\small\tt torch.compile}~\cite{Ansel_2024} or
JAX~\cite{Frostig_2018}  
are required to analyse linear algebra expressions
and select or generate optimised GPU kernels via a just-in-time (JIT) compilation approach.

However, it is possible to obtain high speed and low overhead
while still providing an intuitive, high-level API
that does not burden practitioners with device-specific or vendor-specific miscellany.
Inspired by the widely-used C++ Armadillo CPU linear algebra library~\cite{Sanderson_2025},
which uses template metaprogramming to optimise expressions at \textit{compile-time},
we introduce and describe the Bandicoot GPU linear algebra library
which uses template metaprogramming
to generate specialised and optimised GPU kernels \textit{entirely} at compile-time.
This approach is in contrast to previous C++ based approaches that rely on \textit{run-time}
expression analysis and/or elaborate JIT compilation infrastructure~\cite{Chen_2012,Malcolm_2012}.
Due to the low overhead and the ability to generate highly specialised kernels,
expressions written in Bandicoot can outperform other toolkits,
sometimes by considerable margins.

Bandicoot aims to facilitate easy adaptation of Armadillo based code,
enabling existing CPU-focused computational pipelines to take advantage of GPUs.
Bandicoot supports multiple GPU backends, including CUDA and OpenCL,
with further backends like Vulkan, HIP/ROCm, and Apple Metal under active development.
As the library is not focused simply on numerical linear algebra for machine learning,
support is provided for various decompositions and factorisations,
such as SVD, eigen, Cholesky, etc.

\section{Intuitive and Simple User Interface}
\label{sec:ui}
\vspace{-1ex}

Bandicoot provides a high-level interface for linear algebra
without sacrificing performance.
To maximise the reusability of existing codebases,
the Bandicoot API is modelled to match the API of the
widely-used Armadillo C++ linear algebra toolkit~\cite{Sanderson_2025}.
Armadillo provides an API that closely approximates the API of MATLAB,
which lessens the work involved for a practitioner
when making a transition from slow, high-overhead tools like MATLAB or Mathematica
to fast C++ implementations.
The simplicity of Armadillo's API,
along with its speed~\cite{Psarras_2022},
is directly responsible for its use in numerous well-known downstream packages,
such as the \textit{mlpack} machine learning library~\cite{mlpack2023},
the \textit{ensmallen} numerical optimisation library~\cite{curtin2021ensmallen},
and the popular \textit{RcppArmadillo} project~\cite{eddelbuettel2014rcpparmadillo}
that provides efficient numerical linear algebra to the R language.

\begin{figure}[b!]
\vspace{-2ex}
\centering
\includegraphics[width=\textwidth]{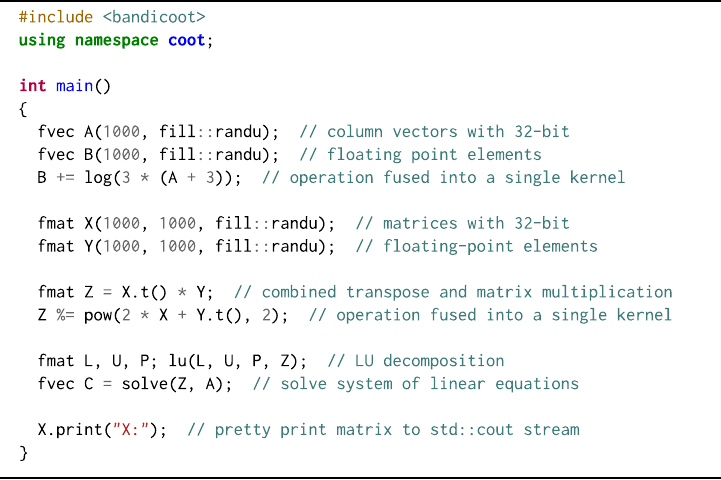}
\vspace{-2ex}
\caption{An example C++ program using Bandicoot to execute linear algebra operations on a GPU.
The corresponding Armadillo-based program, which uses the CPU instead of the GPU,
can be obtained by replacing the code on line 1 with \mbox{\small\tt \#include <armadillo>}
and on line 2 with \mbox{\small\tt using namespace arma;}.}
\label{fig:prog}
\end{figure}

Using Bandicoot does not require deep GPU knowledge;
explicit setup or management of the device is not necessary
unless specifically desired.
An example program is shown in Figure~\ref{fig:prog}.
All operations are dispatched to the GPU asynchronously,
allowing the GPU to compute the results while the main thread of the C++ program continues.
GPU synchronisation is only performed when necessary
(e.g. within the \mbox{\small\tt X.print()} function),
or when explicitly specified via the \mbox{\small\tt coot\_synchronise()} function.

In general, it is possible to take an existing Armadillo program and
replace uses of Armadillo matrices and operations (e.g. \mbox{\small\tt arma::fmat})
directly with Bandicoot matrices and operations (e.g. \mbox{\small\tt coot::fmat}),
resulting in the computations running on the GPU instead of the CPU.
Although not all Armadillo operations are yet implemented in Bandicoot,
work is ongoing for that, including porting factorisations from MAGMA~\cite{Abdelfattah_2024}. 

However, not all Armadillo code will immediately run efficiently on a GPU.
This is due to the different nature of GPU computations.
For example,
individual element access (e.g. \mbox{\small\tt X(r,c) = val}) on a CPU has very low overhead.
In contrast, accessing an element of a Bandicoot matrix incurs a high-latency copy from GPU memory to CPU memory.
Therefore, Bandicoot code is generally much more efficient when written as
a series of matrix-level operations instead of loops over individual elements.
This is similar to the recommendations made by other GPU focused toolkits.

\section{Internal Design and Expression Optimisation}
\label{sec:general_design}
\vspace{-1ex}

\begin{figure}[b!]
\centering
\includegraphics[width=0.84\textwidth]{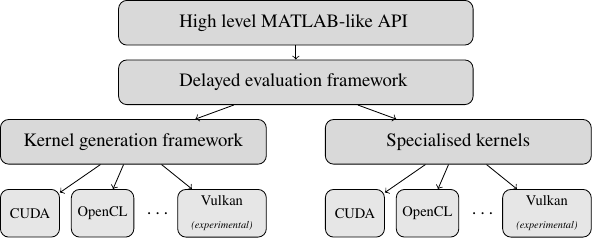}
\vspace*{-1ex}
\caption{Internal architecture and design of Bandicoot.
High-level user
code is converted to efficient, fused GPU kernels at program compilation time,
not at run-time.  Vulkan backend support is currently experimental; more backends (e.g. HIP/ROCm) are~planned.}
\label{fig:arch}
\end{figure}

As Bandicoot is intended to be vendor-independent,
it has been designed so that low-level device implementation is
fully separated from
the user interface code and
code related to optimising linear algebra expressions.
Figure~\ref{fig:arch} outlines the internal design.

Underneath the user-facing functions
(e.g. \mbox{\small\tt operator+()}, \mbox{\small\tt accu()}, \mbox{\small\tt trans()}, etc)
is a template metaprogramming infrastructure
that provides significant compile-time optimisations of mathematical expressions.
This is accomplished by automatically collecting (at compile-time)
the structure of a linear algebra expression as an elaborate custom type,
and then using techniques such as template specialisation~\cite{Vandevoorde_2017}
to generate efficient code to evaluate the expression.

As an example,
the expression \mbox{\small\tt X.t() + 3}
is represented by the C++ type
\mbox{\small\tt eOp<Op<fmat, op\_htrans>, op\_scalar\_add>},
where \mbox{\small\tt op\_htrans} and \mbox{\small\tt op\_scalar\_add} are
placeholder types that represent the operations to be performed.
In essence, the C++ type of an expression is the Abstract Syntax Tree (AST)~\cite{Harper_2016} of that expression,
and the use of template specialisations allows the compiler to optimise that AST.
Furthermore,
the entire expression scaffolding is optimised out of the program at compile-time,
and thus no traces of it remain in the compiled code.
More details on the infrastructure can be found in accompanying literature~\cite{Curtin_2026,Sanderson_2025,sanderson2017armadillo}.

Unique to Bandicoot among linear algebra toolkits is
its compile-time generation of fused GPU kernel sources,
supported by all available GPU backends.
Every GPU backend supports on-the-fly compilation of GPU kernel source,
so that user code can freely be distributed between systems with various GPUs.
Most linear algebra toolkits make use of this,
and ship with either a set of predefined kernel sources
or infrastructure to build these kernels when they are needed.
For specialised operations such as matrix decompositions and pseudo-random number generation,
Bandicoot has predefined kernel sources that are compiled in exactly this way.

However, because Bandicoot has the AST for each expression available at compile-time,
it can generate fused GPU kernels directly at compile-time.
These generated GPU kernel sources can then be embedded directly into the compiled program.
This reduces the size of the compiled program,
as only those kernels that are needed are included,
and no run-time infrastructure for generating kernels is necessary.

\begin{figure}[t!]
\centering
\includegraphics[width=\textwidth]{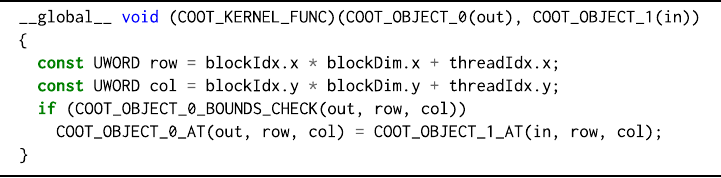}
\vspace{-4ex}
\caption{A simplified version of Bandicoot's \mbox{\small\tt copy} skeleton kernel
for the CUDA backend.
The definition of each \mbox{\small\tt COOT\_\*()} macro is generated at compile-time.}
\label{fig:skel_copy}
\end{figure}

The approach relies on the use of \textit{skeleton kernels}
and C-like macros, which are supported in some form by every GPU vendor's low-level toolkits.
Each skeleton kernel is a deliberately straightforward recipe for a kernel
that uses placeholders to represent its inputs and outputs.
Figure~\ref{fig:skel_copy} shows a
simplified example of Bandicoot's \mbox{\small\tt copy} skeleton kernel,
which copies an input Bandicoot expression to an output.

When a user writes a linear algebra expression,
Bandicoot constructs the AST of that expression as a compile-time type.
Each macro can be constructed by recursing on the AST
and applying individual rules to nodes of the AST
via partial template specialisation.
In the following text we briefly demonstrate this process applied
to the example expression \mbox{\small\tt X.t() + 3},
which yields the type \mbox{\small\tt eOp<Op<fmat, op\_htrans>, op\_scalar\_add>}.

Let us consider the element access macro in the skeleton \mbox{\small\tt copy} kernel:
\mbox{\small\tt COOT\_OBJECT\_1\_AT(out, row, col)}.
For the simple base case type of \mbox{\small\tt fmat} (a floating-point matrix),
we can use the following trivial definition,
where \mbox{\small\tt in\_n\_rows} represents the number of rows in the matrix:

\begin{center}
\noindent \mbox{\small\tt COOT\_OBJECT\_1\_AT(in, row, col) = (in[row + col * in\_n\_rows])}
\end{center}

At the next level up, the type is \mbox{\small\tt Op<fmat, op\_htrans>}
(i.e., a transposed matrix).
We can generate the element access macro for this as the transposed version
of the inner argument of the \mbox{\small\tt Op}'s element access macro,
where the \mbox{\small\tt row} and \mbox{\small\tt col} variables are swapped on the right hand side:

\begin{center}
\noindent \mbox{\small\tt COOT\_OBJECT\_1\_AT(in, row, col) = (in[col + row * in\_n\_rows])}
\end{center}

Finally, the outermost \mbox{\small\tt eOp} is considered.
For this element access macro, we need to add the value of the scalar
to the element access macro of the inner argument of the \mbox{\small\tt eOp}:

\begin{center}
\noindent \mbox{\small\tt COOT\_OBJECT\_1\_AT(in, row, col) = (in[col + row * in\_n\_rows] + aux)}
\end{center}

\noindent where \mbox{\small\tt aux} is the variable representing the scalar value
(in our example, the \mbox{\small\tt 3} in \mbox{\small\tt X.t() + 3}).

This general strategy is used to traverse the AST of any expression and generate
a list of kernel arguments (represented via \mbox{\small\tt COOT\_OBJECT\_x()}),
bounds checks (\mbox{\small\tt COOT\_OBJECT\_x\_BOUNDS\_CHECK()}),
element access (\mbox{\small\tt COOT\_OBJECT\_x\_AT()}),
a unique generated kernel name (\mbox{\small\tt COOT\_KERNEL\_FUNC}),
and a few other macros that are used by various skeleton kernels.

Importantly, this means that each skeleton kernel can be used for
virtually arbitrary expressions.
For instance, the \mbox{\small\tt copy} skeleton kernel
is used for almost every Bandicoot expression of the form
\mbox{\small\tt Y = <rhs>}
for some right-hand-side expression.
This reuse of skeleton kernels significantly reduces the number of GPU kernels
that must be implemented and maintained in Bandicoot,
and compares favourably to other GPU toolkits
that maintain hand-implemented kernels for a variety of situations.

The AST traversal example given above
is a simplified version of the actual implementation in Bandicoot.
The exact details of the macro generation framework code can be found in the
\mbox{\small\tt include/bandicoot\_bits/kernel\_gen/} subdirectory,
present in Bandicoot~4.x versions.
The framework is built around the utility \mbox{\small\tt char\_array<N>} struct,
which is a compile-time string of length \mbox{\small\tt N}
that supports compile-time concatenation with other compile-time strings.

\section{Empirical Speed Comparison}
\label{sec:experiments}
\vspace{-1ex}

To demonstrate the efficacy of Bandicoot's
template metaprogramming approach,
we compare with other toolkits on
a variety of simple and complex linear algebra expressions.
We use the following expressions,
written here using the Bandicoot API,
with capitalised variables representing matrices
and lowercased variables representing scalars.
The {\tt \%} operation represents the Schur product (element-wise multiplication),
which is distinct from the {\tt *} operation that represents matrix multiplication.
For submatrix expressions, we use the center of the matrix with
half the rows and columns.

\begin{enumerate}[label=\textbf{\arabic*.}, ref=\arabic*]

\item \textbf{Basic operations and submatrices:} \vspace{1ex} \\
\begin{tabular}{ll}
{\small \tt add2}: & \mbox{\small\tt Z = X + Y; } \\
{\small \tt add4}: & \mbox{\small\tt Z = A + B + C + D; } \\
{\small \tt chain}: & \mbox{\small\tt Z = A * B * C * D; } with decreasing sizes for each matrix. \\
{\small \tt addsub2}: & same as \mbox{\small\tt add2}, but using submatrices of \mbox{\small\tt A} and \mbox{\small\tt B}. \\
{\small \tt addsub4}:\ \ & same as \mbox{\small\tt add4}, but using submatrices of \mbox{\small\tt A}, \mbox{\small\tt B}, \mbox{\small\tt C}, and \mbox{\small\tt D}. \\
\end{tabular}
\\

\item \textbf{Compound expressions:} \vspace{1ex} \\
\begin{tabular}{ll}
{\small \tt expr1}: & \mbox{\small\tt Z = 2 * (X.t() + Y) + 2 * (X + Y.t()); } \\
{\small \tt expr2}: & \mbox{\small\tt Z = a * A + (B + C).t() + log(pow(D, 2)); } \\
{\small \tt expr3}: & \mbox{\small\tt Z = 1 / (X \% conv\_to<fmat>::from(Y) + log(log(X + 2) \% W)); } \\
{\small \tt diagsum}: & \mbox{\small\tt (X.diag(-1) + X.diag(1)) \% (Y.diag(-1) + Y.diag(1))} \\
\end{tabular}
\\

\item \textbf{Neural network activations:} \vspace{1ex} \\
\begin{tabular}{ll}
{\small \tt relu}: & \mbox{\small\tt Z = X \% (X > 0); } \\
{\small \tt sigmoid}: & \mbox{\small\tt Z = 1 / (1 + exp(-X)); } \\
{\small \tt swish}: & \mbox{\small\tt Z = X / (1 + exp(-beta * X)); } \\
{\small \tt gelu}: & \mbox{\small\tt Z = (X / 2) \% (1 + tanh(sqrt(2 / pi) * (X + alpha * pow(X, 3)))); } \\
\end{tabular}

\end{enumerate}

\clearpage\newpage

\begin{figure}[b!]
\centering
\includegraphics[width=\textwidth]{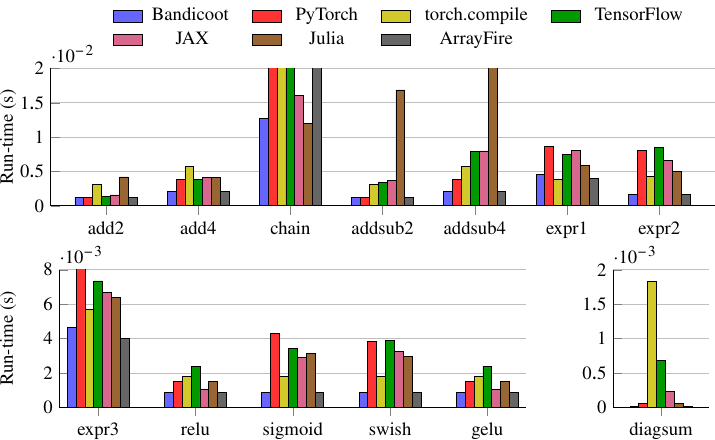}
\vspace{-4ex}
\caption{Average runtime (in seconds) for 50 trials of each expression,
excluding warm-up costs.
Bandicoot is able to fuse operations into a single kernel at compile-time,
resulting in execution speed that is the fastest or on par with the fastest toolkit
in the vast majority of cases.}
\label{fig:res1}
\end{figure}

We ran each of these expressions 50 times,
after two `warmup' runs to handle any JIT compilation and/or run-time costs,
and computed the mean of each expression's run-time.
In all cases, the standard deviation of the run-times was negligible.
Our test setup used an NVIDIA RTX 4090 GPU,
with the CUDA backend used for each toolkit.
Except for the \mbox{\small\tt chain} example, we used
square matrices with $10$k rows/columns
and 32-bit floating point numbers, totalling $\sim\!\!382$ MB per matrix.
Results are shown in Figure~\ref{fig:res1}.

Bandicoot is the fastest or on par with the fastest toolkit for the vast majority of the expressions.
It must be noted that Figure~\ref{fig:res1} focuses only on the time it took to run a kernel
after warmup runs,
and as such the run-times do not include extra overheads such as JIT compilation.
Bandicoot has no such overhead,
beyond the GPU driver's initial compilation of the generated kernel.
Even ArrayFire,
which is used directly from C++,
suffers from some overhead as it has an internal JIT compiler.

Next, to demonstrate the ability of Bandicoot's template metaprogramming framework
to scale to an arbitrary number of matrices at compile-time,
we adapted the \mbox{\small\tt add2} example to add an increasing number of matrices
and computed the throughput achieved (in GB/s)
by each toolkit.
This is a memory-bound operation,
and thus we can easily compute the peak possible bandwidth.
For the RTX~4090, with a 384-bit GDDR6X memory bus clocked at 1313~MHz,
this is theoretically 1008~GB/s.
Peak speeds are generally unachievable in practice;
using NVIDIA's \mbox{\small\tt nvbandwidth} tool,
we computed a maximum achievable bandwidth of 868~GB/s.

\clearpage\newpage

\begin{figure}[t!]
\centering
\includegraphics[width=\textwidth]{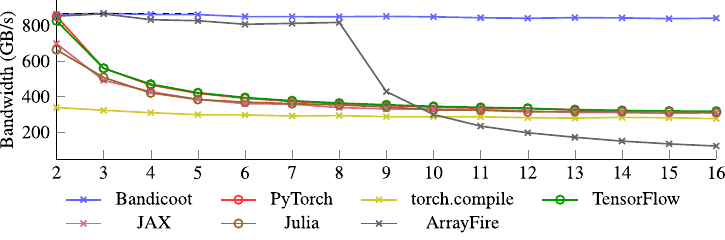}
\vspace*{-4ex}
\caption{Effective throughput of matrix addition as a function of the number of
matrices to be added.  Dashed line indicates peak possible memory bandwidth.}
\label{fig:res2}
\end{figure}

We again used square matrices with $10$k rows/columns,
averaging the run-time over 50 trials.
Figure~\ref{fig:res2}
plots observed throughput as a function of the number of added matrices,
ranging from 2 to 16.
Only Bandicoot is able to maintain maximum achievable bandwidth across all tested cases.
The achievable bandwidth for most other toolkits quickly drops beyond 2 matrices,
and beyond 8 matrices it plateaus at about half the maximum bandwidth.
While the JIT framework in ArrayFire is able to maintain high bandwidth up to 8 matrices,
the throughput severely degrades past that point,
eventually falling well below the other toolkits.

%
%
%

\section{Concluding Remarks}
\label{sec:conclusion}

We have described Bandicoot,
a GPU linear algebra library
implemented in C++.
Its API matches the Armadillo library,
and thus existing user code can be easily ported to make use of GPUs.
Internally, Bandicoot uses an elaborate template metaprogramming framework
that allows the fusion of operations within a compound mathematical expression into a single GPU kernel,
meaning that the expression is computed in a single pass over all given input matrices.
This is achieved at compile-time and hence does not require any JIT infrastructure
or other overheads at run-time.
Comparative evaluations demonstrate competitive performance
and the ability of the generated kernels to saturate memory bandwidth.

The software is licensed under the permissive Apache License, version 2.0,
and can be found at \url{https://coot.sourceforge.io}.
Future work includes adding more backends,
including Apple Metal and AMD HIP/ROCm,
support for low-precision floating point types,
and eventual implementation of all Armadillo functionality.

~

~

\begin{small}
\noindent {\bf Acknowledgements.}
Ryan Curtin's contributions are based
on work supported
by the National Aeronautics and Space Administration (NASA)
under the ROSES-23 HPOSS program, grant no.~80NSSC24K1524.
Ryan Curtin's and Marcus Edel's contributions have also been commissioned and supported by the Sovereign Tech Fund.
\end{small}

\vspace{-1ex}

\clearpage\newpage

\def~{\,}  

\bibliographystyle{splncs04}
\bibliography{refs}

\end{document}